\documentclass[journal]{IEEEtran}

\ifCLASSINFOpdf
\usepackage[pdftex]{graphicx}
\usepackage{longtable}
\usepackage{supertabular}
\usepackage{multicol}
\usepackage{empheq}
\usepackage{amsmath}
\usepackage{cite}
\usepackage{makecell}
\usepackage{color}
\usepackage{cases}
\usepackage{amsmath}
\usepackage{multirow}
\usepackage{xcolor}
\usepackage{breqn}
\usepackage{float}
\usepackage{upgreek}
\usepackage{subfig}

\DeclareGraphicsExtensions{.png}
\else

\fi

\begin{document}
\title{Conceptual Design and Analysis of No-Insulation High-Temperature Superconductor Tubular Wave Energy Converter}
\author{Kyoungmo Koo, Wonseok Jang, Jeonghwan Park, Jaemyung Cha and Seungyong Hahn
\thanks{K. M. Koo, C. Im, G.Kim, J. Kim and S. Hahn are with the Department of Electrical and Computer Engineering, Seoul National University, Seoul 08826, Korea (e-mail: hahnsy@snu.ac.kr).}
}

\maketitle

\begin{abstract}
So far, a number of wave energy converters (WEC) have been proposed to increase efficiency and economic feasibility. Particularly, tubular WEC with permanent magnets and coil winding packs is mostly used to convert the wave energy. Due to the demand for high magnetic flux density in WEC, research has been conducted on high-temperature superconductors (HTS) WEC. In this paper, the conceptual design of no-insulation (NI) HTS tubular WEC and its optimization process are proposed. Using NI technology, it has become possible to design WEC with high volumetric efficiency and cost-effectiveness. Furthermore, the design is analyzed in the aspect of electromagnetism, mechanical force, and cryogen. The performance of the proposed WEC is evaluated as a response to various waveforms and their amplitudes. A rectifying circuit of WEC connected in parallel with load resistance is used for the output power study.

\end{abstract}
\IEEEpeerreviewmaketitle
\section{Introduction}
\IEEEPARstart{T}{he} high demand for electricity stemming from the expansion of the electric mobility industry implies an upcoming electric power shortage problem in the future. Also, there has been growing concern about global warming due to fossil fuel usage. These factors cast light on the importance of renewable energy harvest. Among renewable energies, wave energy has been emphasized as a growing energy resource. Wave energy around the world has been increasing annually by 0.47 \% from 1948 to 2008 \cite{reguero2019recent}. (Southern Ocean 0.58 \%, Pacific 0.35 \%, Atlantic and India 0.26 \% per year) Besides, wave energy is one of the most affluent energy resources around the world. Total wave power around the coastlines of the world is about 2.11 TW. Among the total wave, the power that can be harvested is about 96.6 GW \cite{gunn2012quantifying}. A large amount of energy has not been harvested due to technical difficulties, compared to inland generator systems, such as wind turbine and solar power system. A number of countries are trying to build wave energy converters (WECs) to avoid future energy shortages and to lower their dependencies on fossil fuels. For instance, in 2021, China had been building a 1 MW WEC in Wanshan. Plus, in the same year, India built a 0.9 MW WEC that uses a wave-powered navigated buoy in Chennai, Tamil Nadu \cite {(oes)_author:_(oes)_2022}.
Meanwhile, the range of NI HTS applications on electric machinery has been broadening, recently. By eliminating the turn-to-turn insulation of the HTS pancake coil, much higher operational reliability is achieved, compared to conventional HTS magnets \cite{hahn2010hts}. Due to its strength, research regarding the NI technique has been conducted actively \cite{hahn201945, fazilleau202038, liu2020world, kim2020design, yoon201626}. Also, NI HTS technology has begun to be adopted in the application area where a high magnetic field is needed, including particle accelerators, magnetic resonance imaging (MRI), and electric motors \cite{hahn2011no, hahn2011no2, wang2019electromagnetic, bong2019design}. Considering that existing WEC is having a tradeoff problem regarding magnetic flux density and volume of a magnet, NI HTS will be a valuable solution. Previously, several studies have been conducted in the field of NI HTS rotating machines, including wind power generators \cite{song2014dynamic, kim2017characteristic, song2013hts}.
Meanwhile, the range of NI HTS applications on electric machinery has been broadening, recently. By eliminating the turn-to-turn insulation of the HTS pancake coil, much higher operational reliability is achieved, compared to conventional HTS magnets \cite{hahn2010hts}. Due to its strength, research regarding the NI technique has been conducted actively \cite{hahn201945, fazilleau202038, liu2020world, kim2020design, yoon201626}. Also, NI HTS technology has begun to be adopted in the application area where a high magnetic field is needed, including particle accelerators, magnetic resonance imaging (MRI), and electric motors \cite{hahn2011no, hahn2011no2, wang2019electromagnetic, bong2019design}. Considering that existing WEC is having a tradeoff problem regarding magnetic flux density and volume of a magnet, NI HTS will be a valuable solution. Previously, several studies have been conducted in the field of NI HTS rotating machines, including wind power generators \cite{song2014dynamic, kim2017characteristic, chae2020design, song2013hts}.

This paper focuses on the design and analysis of WEC whose field winding part is substituted with NI HTS magnets. In such a process, magnet design for the minimum material cost was found through the parameter sweep analysis. Also, multiphysics design approaches including electromagnetics, solid mechanics, and cryogenic engineering for the design of NI HTS tubular WEC are conducted. Last but not least, operational analysis has been conducted for several ocean waves with different amplitudes. Output power for the design was calculated through a three-phase bridge rectifier circuit including its load resistance.

\section{SYSTEM CONFIGURATION}
\begin{figure} [h]
\centering
\includegraphics[width=1.0\linewidth]{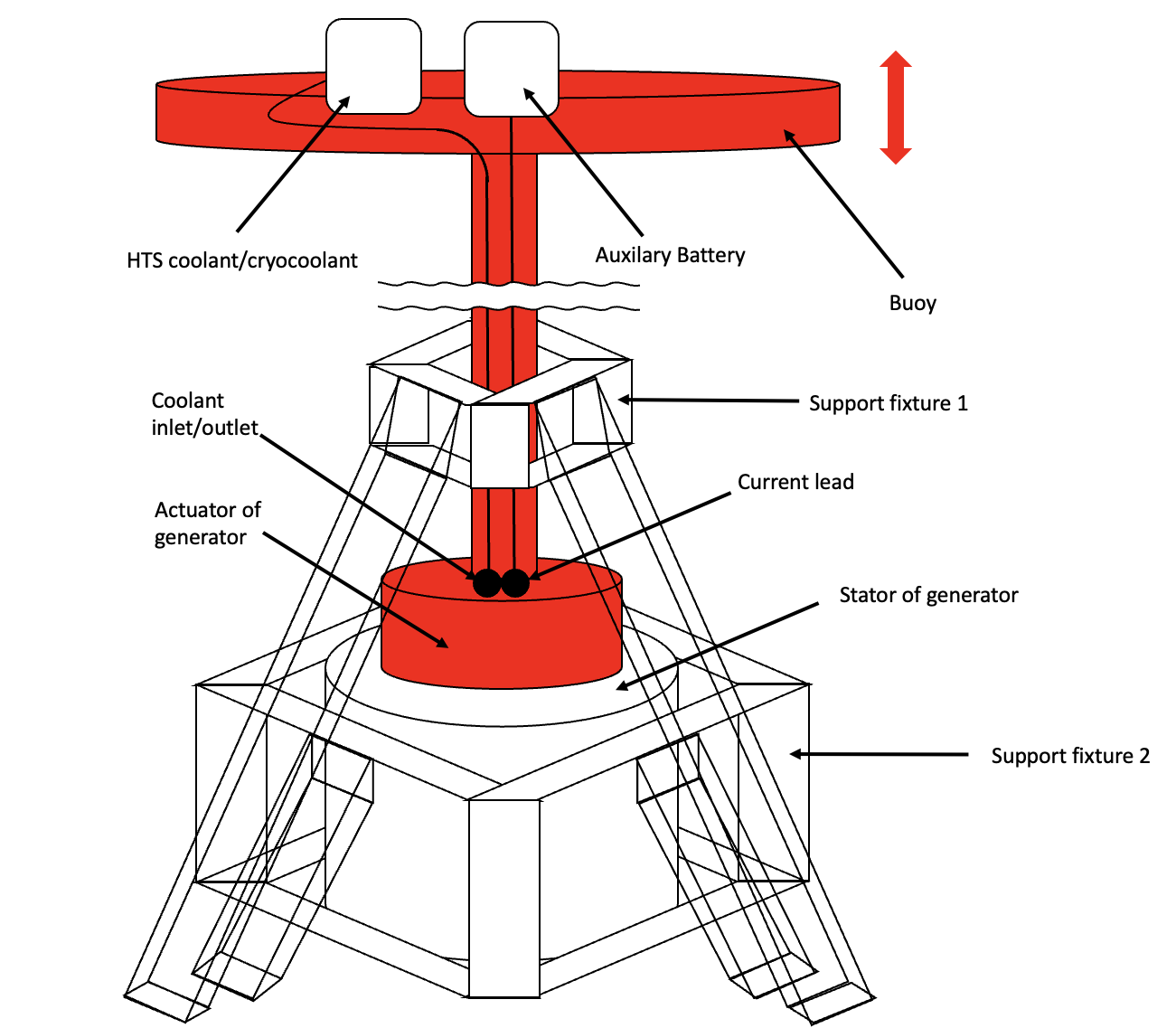}
\centering
\caption{Conceptual design of WEC system.}
\label{fig: WEC_system}
\end{figure}

Point absorber is chosen among various WEC designs, which uses oscillation of buoy to extract wave energy and drive generator \cite{faizal2014design}. A point absorber is a typical WEC structure since it can generate power from any directional wave flow. Fig.  \ref{fig: WEC_system} illustrates how the buoy moves according to the wave and moves the actuator to generate electricity.

As the buoy fluctuates according to an ocean wave, the actuator part of the generator that is connected to the buoy also moves and the voltage is induced in the coil by Faraday's law. It should be noted that the figure is not a exactly scaled version of the actual system. Connecting part between the buoy is about 30 m, which means that its actual length is much longer than the figure. The basic structure of WEC is mainly derived from \cite{shmilovitz2005definition}. Cryocooler for HTS cooling and auxiliary battery which is used for constant current flow in HTS is placed on a buoy. The wire for those components is connected to HTS through the structure between the buoy and the generator. The power that the cryocooler consumes come from a generator and the power is rectified by a converter that will be discussed in 5.1.

\begin{figure} [h]
\centering
\includegraphics[width=1.0\linewidth]{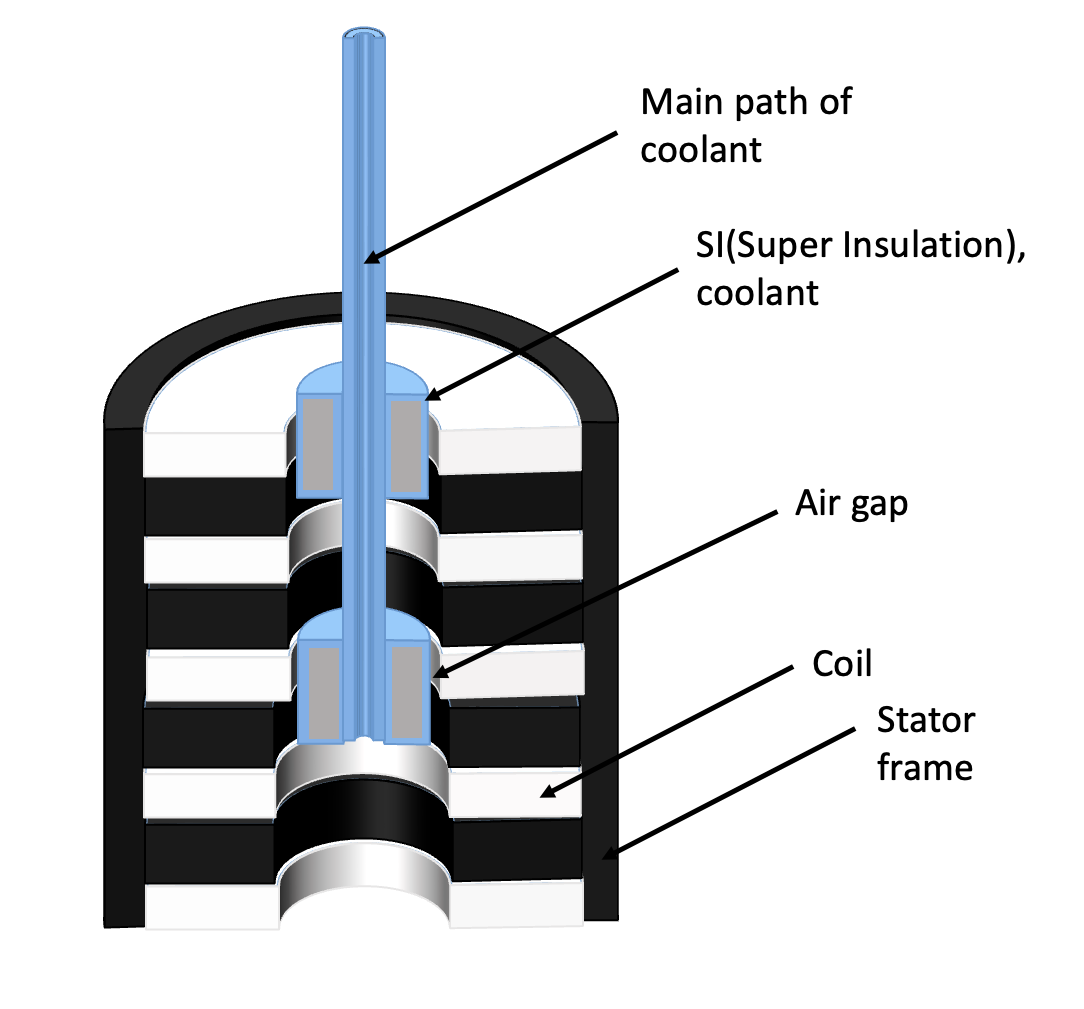}
\centering
\caption{Stator and actuator of WEC generator.}
\label{fig: WEC_generator}
\end{figure}

LH$_2$ is used as the coolant of the generator. Since LH$_2$ initially enters the actuator, it will flow through its path until it arrives at the lowest HTS magnet among DP (double pancake). Dunking is used as a cooling method, which fully immerses magnets into the coolant. Fig. \ref{fig: WEC_generator} shows the specific structure of the generator and cooling mechanism for two HTS magnets. Coolant follows the main path and enters each magnet. Super insulation surrounds each HTS magnet and protects it from transferred heat. The actuator frame and stator frame are both made of fiber-reinforced plastic (FRP) to reduce total mass, except for the outer yoke of the stator frame, which is iron. By using NI HTS technology, the high magnetic flux density can be increased effectively in terms of volume. The mass of the magnet is also greatly reduced, leading to the enhancement of mechanical stability. As typical generators, an air gap between the actuator and stator will be maintained using bearings. Every neighboring HTS magnet will have an opposite magnetic pole for the dramatic rate of magnetic flux change per second. The phase of each coil is designed to be apart from each other, composing a three-phase circuit. A rectangular form coil is used for the armature part and SuNAM superconducting tape is used for HTS magnets.
Table \ref{Key Operational Parameters of WEC} shows key parameters of the design of WEC and ocean waveform during operation.
\begin{table}[h]
\renewcommand{\arraystretch}{1.5}
\caption{Key Operational Parameters of WEC}
\label{Key Operational Parameters of WEC}
\centering
\begin{tabular}{c c c}
\hline
\hline
\bfseries Parameters &  \bfseries Units & \bfseries Value  \\
\hline
Buoy radius & [m] & 0.510\\
Buoy height  & [m] & 0.300\\
Number of magnets & [-] & 4\\
Number of slots & [-] & 12 \\
Bobbins per phase & [-] & 4\\
Magnet height & [mm] & 119.80\\
Magnet width & [mm] & 22.50\\
Air gap & [mm] & 10\\
Cryo-vessel height & [mm] & 131.80  \\
Cryo-vessel width & [mm] & 34.50 \\
Armature coil height & [mm] & 50 \\
Armature coil width & [mm] & 220 \\
Actuator height & [mm] & 1150 \\
Actuator radius & [mm] & 100\\ 
Total height & [mm] & 1150 \\
Total radius & [mm] & 400 \\
Operating current & [A] & 179 \\
Operating temperature & [K] & 20\\
Coolant & [-] & LH$_2$ \\
Total HTS usage 4 mm; 6 mm & [km] & 3.57; 2.38 \\
Waveform & [-] & Sinusoidal \\
Wave amplitude & [m] & 1.25 \\
Frequency & [Hz] & 0.167 \\
\hline
\hline
\end{tabular}\\
\end{table}

\section{DESIGN APPROACH}
\subsection{Overall design approach and considering ocean wave condition}
A variety of parameters had to be considered during the design. The number of slots per pole is simply designed as three since the THD (Total Harmonic Distortion) and performance remained the same while multiplying the number. THD is calculated as the sum of the energy of all frequencies except fundamental frequency, divided by that of fundamental frequency \cite{jing2018electromechanical}. There have been some significant changes compared to previous works. Compared to the design of \cite{shmilovitz2005definition}, the cross-sectional surface of HTS has been reduced by 40 \%. The air gap has increased by 25 \% for more mechanical safety.
Buoy diameter is determined regarding resonance. Assuming the wavelength of the ocean wave is longer than four times of buoy diameter, buoy diameter is 1.5\% of wavelength and absorbs 99~100 \% of maximum energy, which is why the buoy radius is designed as 0.510 m, assuming the wavelength of 34 m \cite{nazari2013design}. The buoy height is designed at 0.300 m, with a 50 \% margin from being completely submerged.
As illustrated in \cite{liu2019linear}, which conducted HTS WEC experiment based on Chinese sea conditions, an actuator is assumed to move at 0.4 m/s. Furthermore, because the average coastline ocean wave period is less than 8 seconds \cite{hughes2016coastal}, the period is supposed to be 6 seconds and the amplitude to be 1.25 m.
\subsection{Optimization: Parameter sweep and multi-width winding method}
\begin{figure} [h]
\centering
\includegraphics[width=1.0\linewidth]{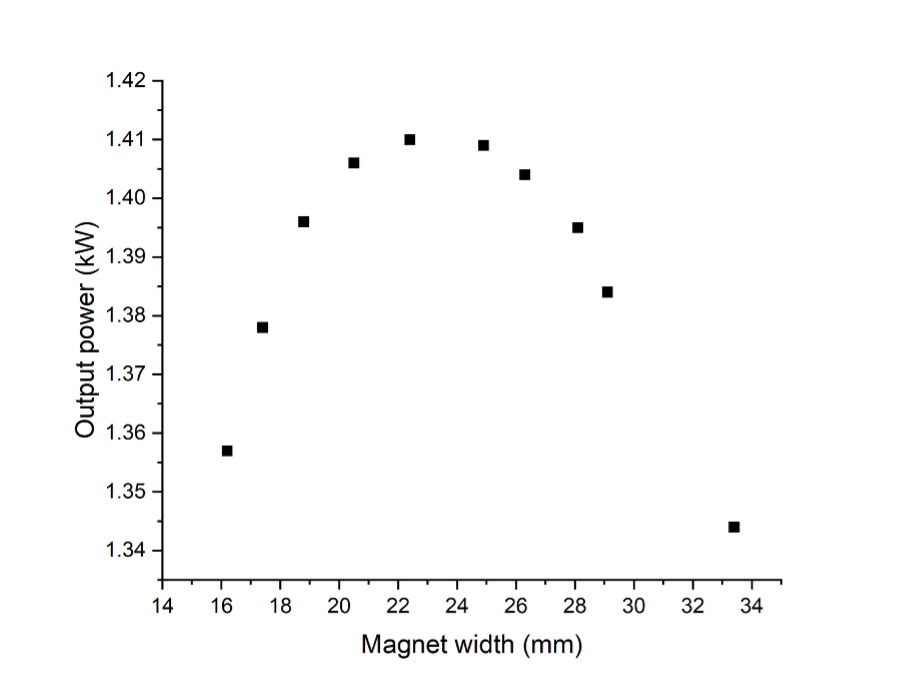}
\centering
\caption{Stator and actuator of WEC generator.}
\label{fig: parameter optimization}
\end{figure}

\begin{figure} [h]
\centering
\includegraphics[width=1.0\linewidth]{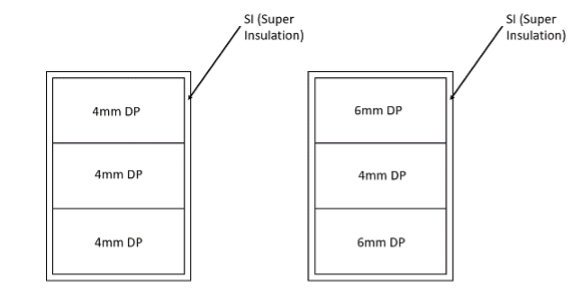}
\centering
\caption{Stator and actuator of WEC generator.}
\label{fig: multi_width_cross_section}
\end{figure}

The system consists of 4 sets of a three-phase circuit, which is composed of 12 coils and 4 HTS magnets in total. A generator is designed to produce target power when connected to the rectifying circuit.
For magnet width and height, parameters that produce the highest voltage using the same total HTS length are determined by the parameter sweep method. As a result, as demonstrated in Fig. \ref{fig: parameter optimization}, magnet height of 119.8 mm, and magnet width of 22.4 mm turned out to produce the highest output power, 14.68 kW.
Additionally, the multi-width method is used to increase critical current, and check whether the output power increases, compared to the single-width model \cite{hahn2013no}. Fig. \ref{fig: multi_width_cross_section} shows the initial design of HTS that is composed solely of 4 mm tape and the multi-width design is made of 6 mm and 4 mm tape. Using the multi-width method, the critical current turned out to be 215 A, which is about 43.3\% higher than the initial design. Also, the multi-width design produced an output power of 14.10 kW, which is about 27.5\% higher than that of the initial design.

\begin{figure}[h] 
    \centering
    \subfloat[single width]{%
        \includegraphics[width=0.2\textwidth]{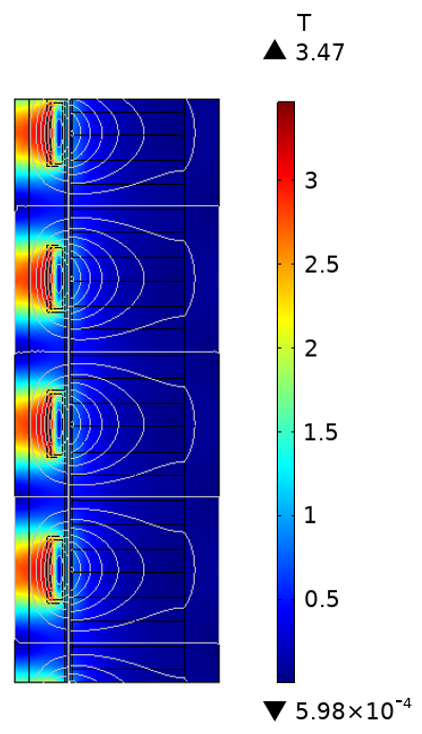}%
        \label{fig:multiwidth_flux_single}%
        }%
    \hfill
    \subfloat[multi width]{%
        \includegraphics[width=0.2\textwidth]{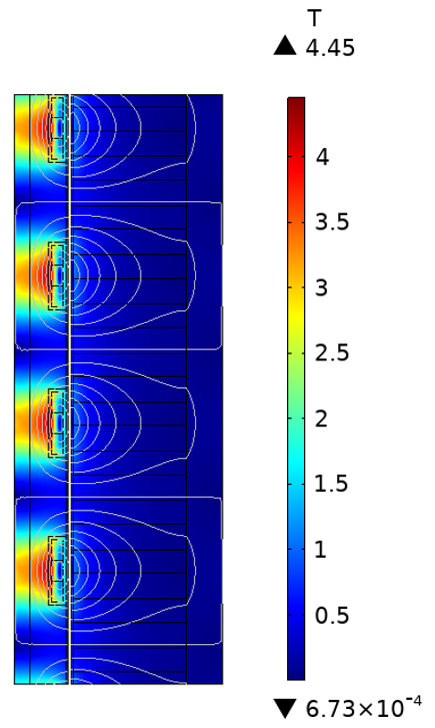}%
        \label{fig:multiwidth_flux_multi}%
        }%
    \caption{Distribution of magnetic flux density in each design: (a) single width, (b) multi width}
\end{figure}

Fig. \ref{fig:multiwidth_flux_single} and Fig. \ref{fig:multiwidth_flux_multi} shows the magnetic flux density norm of each model. As shown in the figure, the multi-width model has a higher maximum magnetic flux density and higher center flux density (4.45 T, 3.14 T), compared to the single-width model (3.46 T, 2.75 T), in the condition of the same conduction current. The result shows that the multi-width method certainly enhanced the magnetic field in the design. Since the performance significantly improved, the design is based on the multi-width model with 4 mm and 6 mm DP.

\subsection{Specific design of HTS magnet}
\begin{figure} [h]
\centering
\includegraphics[width=1.0\linewidth]{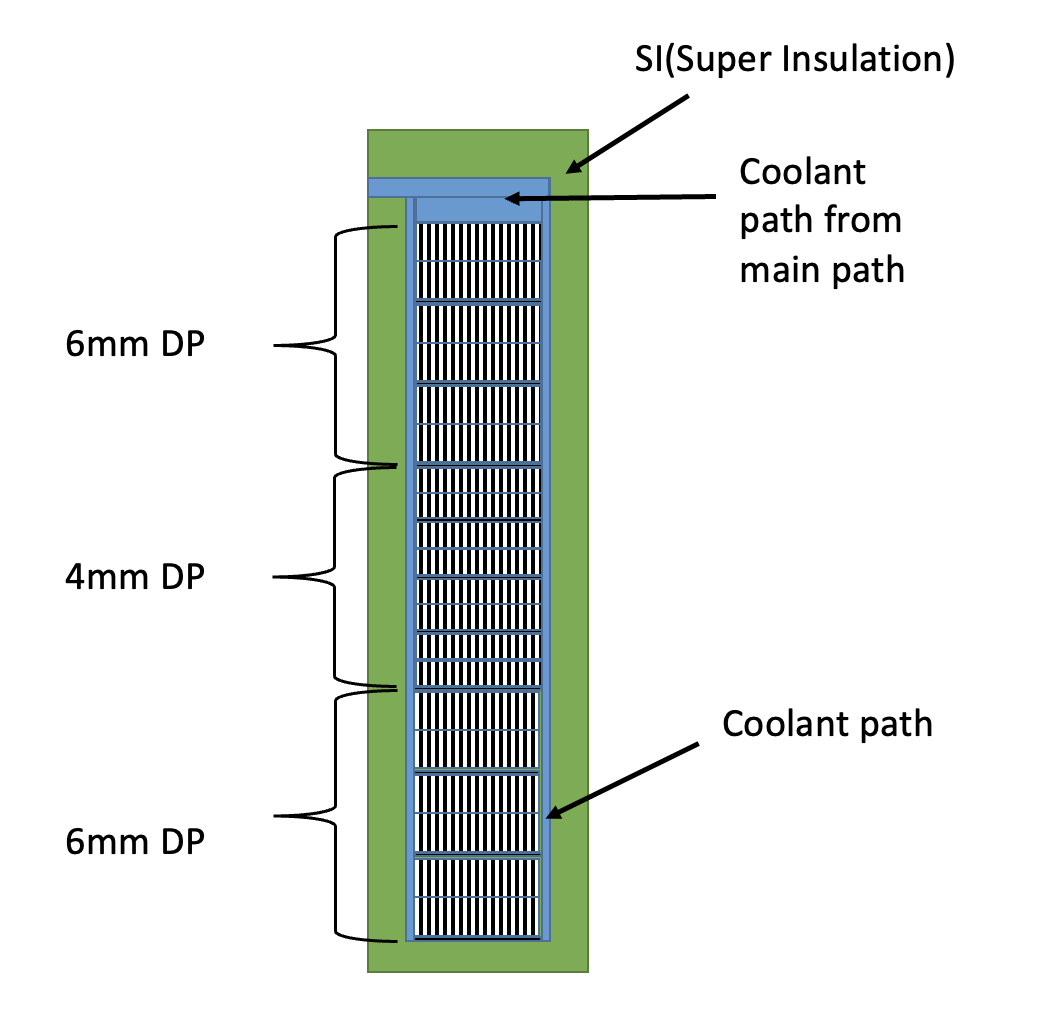}
\centering
\caption{ Cross-sectional view of HTS magnet and insulation}
\label{fig: HTS_cross_section}
\end{figure}

The design of the HTS magnet is as follows. The magnet is composed of a winding pack, insulation plate, and coolant for maintaining an operating temperature of 20 K. HTS magnet is composed of 13 layers as shown in Fig. \ref{fig: HTS_cross_section}. All layers are composed of one 4 mm DP bulk or one 6 mm DP bulk. Additionally, 0.5 mm insulation plates are placed between two DPs. Each DP is 22.4 mm in width, letting the coolant flow on both sides. Also, the upper path, which has a width of 6mm enables effective cooling of the whole magnet. As shown in the figure, coolant flows from the main path, which starts from the top of the actuator. SI (Super Insulation) is designed to have a 6 mm width on each side, making a total magnet height of 131.8 mm and width of 34.5 mm.

\section{Analysis}
\subsection{Electromagnetic analysis: magnetic field and critical current of HTS magnet}
\begin{figure} [h]
\centering
\includegraphics[width=1.0\linewidth]{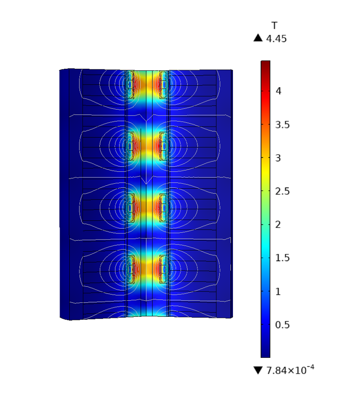}
\centering
\caption{Distribution of magnetic field in the generator}
\label{fig: generator_flux}
\end{figure}

\begin{figure}[h] 
    \centering
    \subfloat[magnetic flux density]{%
        \includegraphics[width=0.2\textwidth]{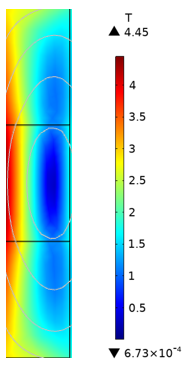}%
        \label{fig:HTS_flux}%
        }%
    \subfloat[critical current of 6 mm DP]{%
        \includegraphics[width=0.2\textwidth]{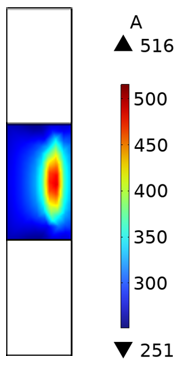}%
        \label{fig:HTS_4mm_current}%
        }%

    \subfloat[critical current of 4 mm DP]{%
        \includegraphics[width=0.2\textwidth]{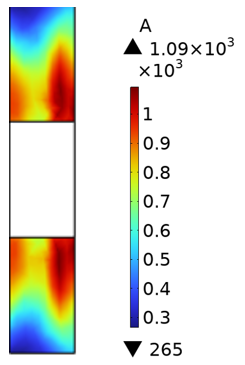}%
        \label{fig:HTS_6mm_current}%
        }%
    \caption{Distribution of critical current and magnetic flux density in HTS magnet ( a)magnetic flux density, b) critical current of 6 mm DP, c) critical current of 4 mm DP)}
\end{figure}

Fig. \ref{fig: generator_flux} shows the distribution of magnetic flux density of the system. As shown in the figure, the maximum magnetic flux density turns out to be 4.5 T in the operating parameters, and the direction of magnetic flux density is mainly the moving direction of the actuator. The left figure shows the distribution.
Fig. \ref{fig:HTS_flux}, \ref{fig:HTS_4mm_current}, \ref{fig:HTS_6mm_current} shows the distribution of critical current and magnetic flux density in a single HTS magnet. As expected, since the magnetic flux density increases as the point of measurement moves toward the center, the critical current turns out to have a tendency of decreasing accordingly. But it is not always the case because the direction of magnetic flux greatly affects the value of critical current at the measured point. Maximum and minimum critical current turned out to be 951.0 A, and 248.9 A at the operating current.

\begin{figure} [h]
\centering
\includegraphics[width=1.0\linewidth]{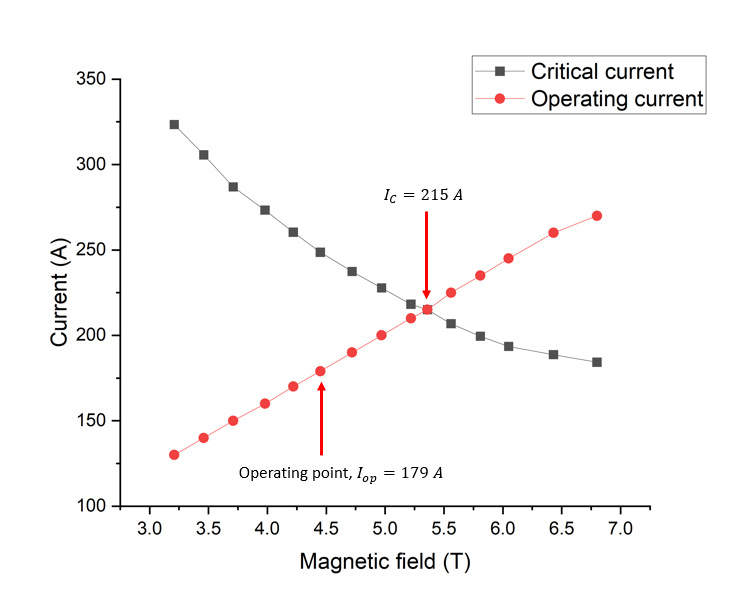}
\centering
\caption{Load line graph of the system}
\label{fig: Load_line}
\end{figure}

Fig. \ref{fig: Load_line} shows the load line graph of critical current. As illustrated in the figure, the system has a critical current value of 215 A, so the current margin obtained by the operating current, 179 A, is calculated as 36 A. The current margin is 20 \% of the operating current, which means that the system is stable in the view of electromagnetism.

Also, to avoid the saturation of the magnetic field that could lead to degradation of system performance, it should be confirmed that the magnetic flux density of the stator yoke should not exceed 1.5 T. It is because the stator yoke is composed of an electrical steel sheet. Fig. \ref{fig: generator_flux} shows the magnetic flux density of the cut plane of the stator and the maximum flux density turns out to be 1.22 T, which means that the stator is free from saturation issues.
\begin{figure}[h] 
    \centering
    \subfloat[Cut plane of stator]{%
        \includegraphics[width=0.2\textwidth]{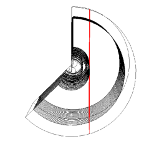}%
        \label{fig:Cut_plane}%
        }%
    \hfill
    \subfloat[Magnetic flux density of stator]{%
        \includegraphics[width=0.2\textwidth]{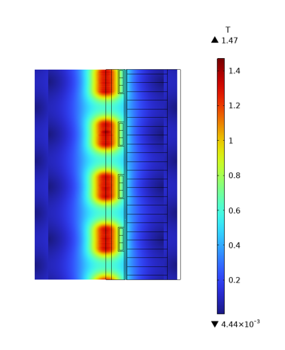}%
        \label{fig:Stator_flux}%
        }%
    \caption{Distribution of critical current and magnetic flux density in HTS magnet ( a) magnetic flux density, b) critical current of 6 mm DP, c) critical current of 4 mm DP)}
\end{figure}
\subsection{Mechanical stability: Lorentz force}
\begin{figure}[h] 
    \centering
    \subfloat[radial stress of HTS magnet]{%
        \includegraphics[width=0.2\textwidth]{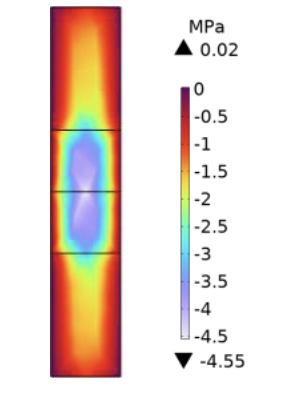}%
        \label{fig:HTS_radial_stress}%
        }%
    \hfill%
    \subfloat[hoop stress of HTS magnet]{%
        \includegraphics[width=0.2\textwidth]{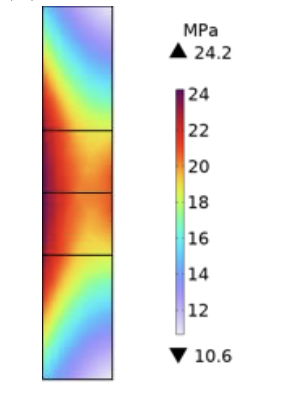}%
        \label{fig:HTS_hoop_stress}%
        }%
    \caption{Distribution of stress of each HTS magnet: (a) radial (b) hoop}
\end{figure}

The following figure is the analysis regarding magnetic force load, due to Lorentz force inside HTS magnets. Fig. \ref{fig:HTS_radial_stress}, \ref{fig:HTS_hoop_stress} shows the distribution of radial and hoop stress, which is created by the inner Lorentz force of the magnet, among four HTS magnets and the maximum value is 24.2 MPa. The distribution of each magnet turned out to be identical because of the symmetry of the design. The design is stable compared to conventional generators, due to the compactness of magnets.

\subsection{ Cryogenic analysis}
Since the design use dunk mode for cooling the magnets,
the initially needed amount of coolant is important for safely operating a generator. The coolant needed to force the temperature of HTS magnets to 20 K is calculated as eq. (\ref{Coolant amount}).

\begin{align}
\label{Coolant amount}
\frac{M_H}{H_{mg}}&=\frac{{h_{Cu}}(300K)-{h_{Cu}}(20K)}{h_H},
\end{align}

M$_H$ is the mass of needed liquid hydrogen for initiating
the generator. M$_{mg}$is the total mass of the magnet, h$_{Cu}$ is the specific heat of copper, and h${_H}$ is the specific heat of hydrogen. The value is calculated as 8.23 kg.
Additionally, the armature coil part should not flow a current of more than 6 A/mm$^2$, in the case of the water- cooled type of design. This will be specifically discussed in 5.2.
Lastly, the critical current of the superconducting wire according to the temperature can be calculated with linear approximation applied as eq. (\ref{T_Ic}), considering the critical temperature of 0 A is 92 K. I$_C$ is a critical current at a certain temperature T. T$_C$ is a critical temperature. I$_{C0}$ is critical current at operating temperature, T$_{op}$.

\begin{align}
\label{T_Ic}
I_C(T)&=I_{C0}\frac{T_C-T}{T_C-T_{op}}, T_{op} < T < T_C
\end{align}

\begin{figure} [h]
\centering
\includegraphics[width=1.0\linewidth]{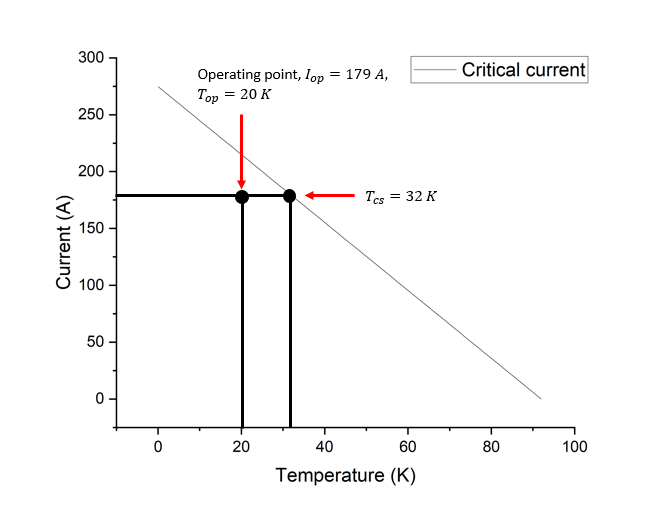}
\centering
\caption{Critical current - temperature graph of HTS magnets}
\label{fig: Critical_current_temperature}
\end{figure}

Using the formula, the stability of the coolant system can be determined by temperature margin. T-I$_C$ graph of SuNAM REBCO tape can be drawn as Fig. \ref{fig: Critical_current_temperature}.

The stability margin of the magnet in terms of enthalpy is also calculated by heat capacity per unit volume, C(T), integrated over temperature margin.

\begin{align}
\label{ER}
E&=\int_{T_CS}^{T_op} C(T) \, dT \,,
\end{align}

Using the upper formula, 8.31 GJ/m$^3$ of energy is
tolerable in the perspective of cryostability.
TABLE \ref{Multiphysics Design Parameters of WEC} shows various parameters that summarize
what is discussed in this section.

\begin{table}[h]
\renewcommand{\arraystretch}{1.5}
\caption{Multiphysics Design Parameters of WEC}
\label{Multiphysics Design Parameters of WEC}
\centering
\begin{tabular}{c c c}
\hline
\hline
\bfseries Parameters &  \bfseries Units & \bfseries Value  \\
\hline
Maximum magnetic flux density & [T] & 4.45\\
Average magnetic flux density  & [T] & 1.94\\
Critical current & [A] & 215\\
Current margin & [A] & 36 \\
\hline
Maximum stress & [MPa] & 24.2\\
Average stress & [MPa] & 11.9\\
Young's modulus of coil & [GPa] & 160\\
\hline
Mass of LN$_2$ & [kg] & 10.17  \\
Critical temperature & [K] & 32 \\
Stability margin & [GJ/m$^3$] & 8.31 \\
\hline
\hline
\end{tabular}\\
\end{table}

\section{OPERATION ANALYSIS}
\subsection{Modeling of the power grid}
\begin{figure} [h]
\centering
\includegraphics[width=1.0\linewidth]{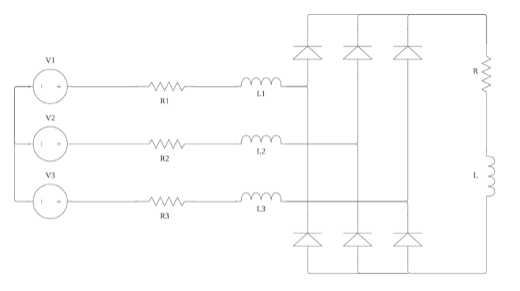}
\centering
\caption{Circuit diagram of the whole system}
\label{Fig: Lumped_circuit}
\end{figure}

Three phase rectifier is connected to the three-phase
voltage sources. 0.15 H of an inductor is connected in series with source and resistance. A resistor is optimized to 4.2 $\Omega$ so as to produce maximum output power. 2.7 H inductor is connected in series with resistance to smooth out the output voltage graph and reduce the distortion rate. Fig. \ref{Fig: Lumped_circuit} shows how the input voltage is rectified through a circuit. The total distortion ratio has greatly decreased after the rectifier from 2.99 to 1.12. The fundamental frequency of the input voltage source turned out to be 0.35 Hz, and that of the output voltage turned out to be 2.01 Hz.

\subsection{Voltage and current analysis of each phase}
\begin{figure} [h]
\centering
\includegraphics[width=1.0\linewidth]{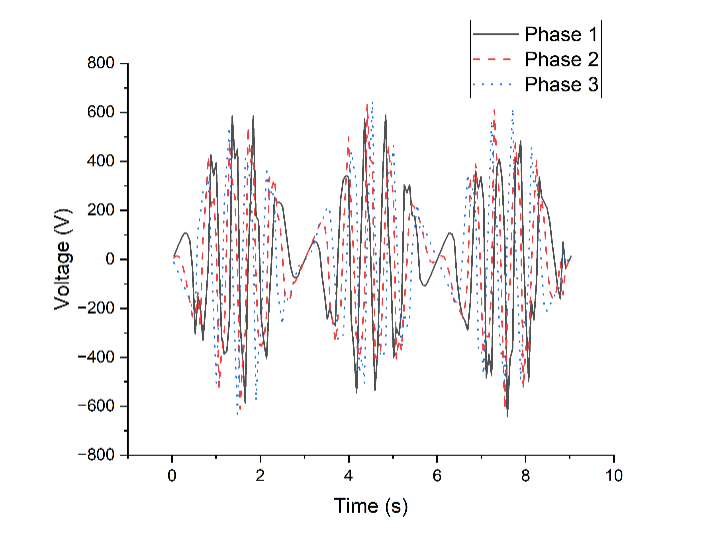}
\centering
\caption{Output voltage of each phase}
\label{fig: Output_voltage_phase}
\end{figure}

Fig. \ref{fig: Output_voltage_phase} shows the voltage results of the design. The voltage shows three cycles of output power waves with changing amplitude as time pasts. The voltage of each phase had a maximum voltage of 589.4 V, 635.0 V, and 646.5 V. V$_{rms}$ (Root Mean Square) of each voltage source turned out to be 267.1 V, 267.2 V, and 264.8 V.

\begin{figure} [h]
\centering
\includegraphics[width=1.0\linewidth]{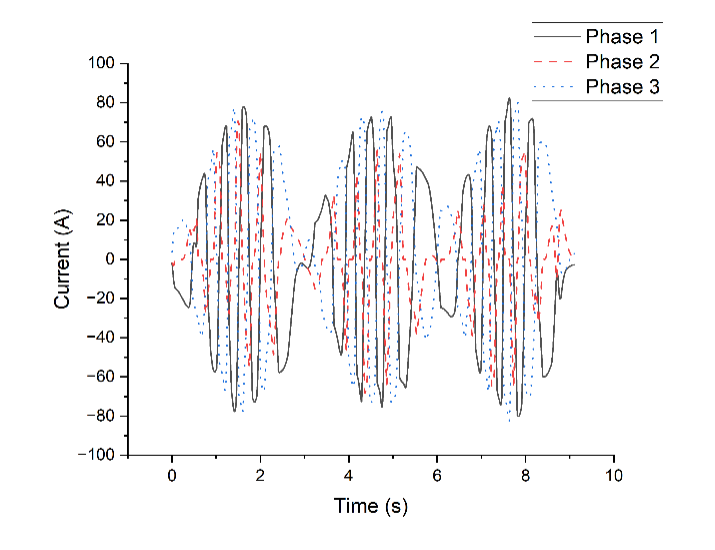}
\centering
\caption{Armature coil current of each phase}
\label{fig: Armature_coil_phase}
\end{figure}

Fig. \ref{fig: Armature_coil_phase} shows the current that flows in each armature coil during a simulation. As stated in 4.3, the current density in the armature coil should be maintained under 6 A/mm$^2$ most time. It turns out that the current in each phase stays under 80 A most time except for several peaks, which are in the range of 80-85 A. Therefore, the system is proven stable considering the cross-sectional surface of the rectangular form coil, considering that the critical current of the armature coil is 140 A.

\subsection{Output power of various conditions}
\begin{figure}[h]
\centering
\includegraphics[width=1.0\linewidth]{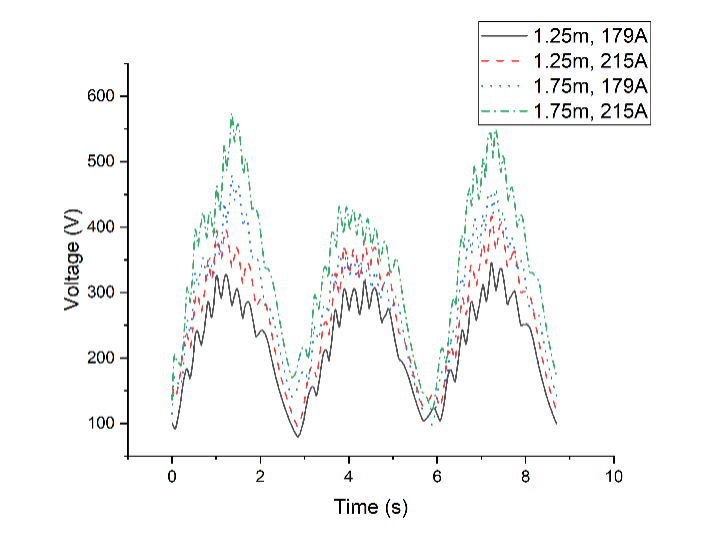}
\centering
\caption{Output voltage of various conditions}
\label{Fig: Output_voltage_conditions}
\end{figure}

Various simulations are conducted with different
conduction currents and wave amplitudes. Fig. \ref{Fig: Output_voltage_conditions} and TABLE \ref{Simulation results of various conditions} show the results of each case. Obviously, voltage wave amplitude was proportional to the conduction current of magnet and wave amplitude, aligning with the result of \cite{ba2016conceptual}. The power, Vrms of input voltage and output voltage, loss, and efficiency of each case is stated in the following table. Since the maximum instant current that flows through the armature coil is maintained under 6 A/mm$^2$ for the 2$^{nd}$, 3$^{rd}$, and 4$^{th}$ cases, the armature coil is stable considering the cooling method.

\begin{figure}[h]
\centering
\includegraphics[width=1.0\linewidth]{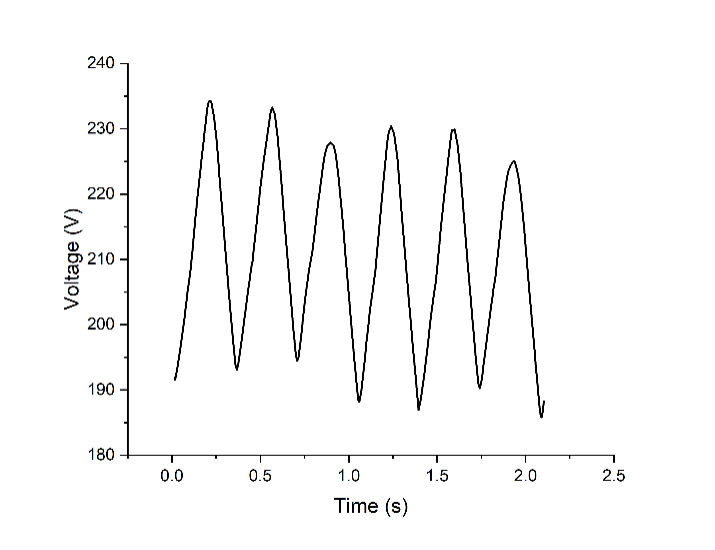}
\centering
\caption{Output voltage of the triangular wave}
\label{Fig: Output_voltage_triangular}
\end{figure}

\begin{table}[h]
\renewcommand{\arraystretch}{1.5}
\caption{Simulation results of various conditions}
\label{Simulation results of various conditions}
\centering
\begin{tabular}{p{0.15\textwidth}p{0.05\textwidth}p{0.05\textwidth}p{0.05\textwidth}p{0.05\textwidth}}
\hline
\hline
\bfseries Wave amplitude, conduction current &  \bfseries 1.25 m, 179 A & 1.25 m, 215 A & \bfseries 1.75 m, 179 A & \bfseries 1.75 m, 215 A  \\
\hline
V$_{rms}$ of input voltage (V) & 284.6 & 347.3 & 462.3 & 555.2\\
V$_{rms}$ of output voltage (V) & 245.8 & 297.6 & 321.5 & 387.5\\
Power (kW) & 14.1 & 21.1 & 24.6 & 35.7\\
Joule loss (kW) & 8.4 & 12.2 & 14.6 & 21.3\\
Efficiency (\%) & 63.0 & 63.4 & 62.7 & 62.7\\
Power factor & 0.84 & 0.84 & 0.84 & 0.84\\
\hline
\hline
\end{tabular}\\
\end{table}
The wave can be also modeled in triangular form. Fig. \ref{Fig: Output_voltage_triangular} compares the output voltage result of triangular form wave and sinusoidal wave. It turned out to be that the case a triangular wave can generate power of 10.63 kW, which is 24.77 \% lower than that of the sinusoidal wave, assuming the same wave amplitude and frequency. The voltage waveform has much less noise and has the form of a sinusoidal wave with a DC offset.

\section{CONCLUSION}
The conceptual design of a no-insulation HTS tubular wave energy generator is proposed in this paper. A multiphysics design approach including electromagnetics, solid mechanics, and cryogenic engineering is conducted for the design. The magnet operates at 20 K, 179 A with the maximum magnetic flux density of 4.45 T.
Based on the proposed design, the voltage results of various waveforms have been obtained. The whole system is modeled as an equivalent circuit and analyzed in terms of the current that flows through armature coils. Using the circuit model, the output power has been calculated for different cases.
Lastly, in the condition of a sinusoidal form ocean wave with an amplitude of 1.25 m and 6 Hz, the output power was calculated as 14.10 kW. The performance turns out to increase as the current inside the magnet and amplitude of the ocean wave increase.
The design shows high volumetric efficiency of electricity production, by applying NI HTS technology. Also, although the paper only discusses the generator production of 10-50 kW, due to the simplicity of the design, production can be increased to a 100 kW-1 MW scale. However, for the scaled-up version, other cooling methods may be needed for the armature part such as forced water cooling, which allows 12 A/mm$^{2}$ \cite{kim2013electromagnetic}, since a much higher current will flow.

\bibliographystyle{IEEEtran}
\bibliography{Ref}

\end{document}